\documentclass[%
 aip,
 amsmath,amssymb,
reprint,%
]{revtex4-1}

\usepackage{graphicx}
\usepackage[caption=false]{subfig}
\usepackage{dcolumn}
\usepackage{bm}

\usepackage[utf8]{inputenc}
\usepackage[T1]{fontenc}
\usepackage{mathptmx}
\usepackage{etoolbox}
\usepackage{xcolor}

\makeatletter
\def\@email#1#2{%
 \endgroup
 \patchcmd{\titleblock@produce}
  {\frontmatter@RRAPformat}
  {\frontmatter@RRAPformat{\produce@RRAP{*#1\href{mailto:#2}{#2}}}\frontmatter@RRAPformat}
  {}{}
}%
\makeatother
\begin{document}

\preprint{AIP/123-QED}

\title[MACOR glass-ceramic based UHV cell for quantum technology applications]{MACOR glass-ceramic based UHV cell for quantum technology applications}
\author{M. Proske}
\email{mproske@uni-mainz.de}
\author{S. Boles-Herresthal}%
\author{D. Latorre-Bastidas}
\author{I. Varma}
\author{R. Skanda}
\affiliation{ 
Institute of Physics, Johannes Gutenberg University Mainz, D-55122, Germany
}%
\author{O. Hellmig}
\author{K. Sengstock}
\affiliation{%
Institute for Quantum Physics, Universität Hamburg, D-20148, Germany
}%
\author{A. Wenzlawski}
\author{P. Windpassinger}
\affiliation{ 
Institute of Physics, Johannes Gutenberg University Mainz, D-55122, Germany
}%

\date{\today}

\begin{abstract}
Compact, customizable, non-magnetic vacuum systems are a key requirement for many field applications of quantum technology based on cold atoms.
We report on the development and construction of a compact, low-cost ultra-high vacuum compatible cell using the glass-ceramic MACOR. The cell offers a CF flange connection to commercial vacuum technology, as well as high numerical aperture viewports for precision optical measurements.
The presented technology shows stable vacuum pressures of $< 1 \cdot 10^{-10}$ mbar for more than a year since the implementation into the vacuum system of a quantum gas experiment, further proving suitability for general quantum technology applications.
\end{abstract}

\maketitle

Precise manipulation of long-lived atomic states, void of any external or thermal perturbation, is a key requirement for most modern quantum technology applications \cite{Schleich2016,Kaltenbaek2021,RevModPhys.94.041001,O'Brien2009,doi:10.1073/pnas.1419326112}. While laser cooling can already minimize the effects of thermal dephasing and decoherence, any further environmental factors are usually dealt with by placing the experimental subject under ultra-high vacuum conditions. These vacuum systems usually consist of metal-based chambers (made from e.g. titanium, aluminium or stainless steel), where glass windows are used as viewports.
While this arguably being one of the cheaper options, the bulky nature of metal to metal connections limits the optical access, especially in applications like optical tweezers and microscopy, where the numerical aperture of optical systems needs to be as high as possible to achieve sub-$\mu$m precision \cite{Pesce2020}. Additionally, the magnitude of parasitic eddy currents on the steel surface is often hard to predict and rather complicated to completely eliminate.
The development of UHV-suitable bonding methods between glass and metal made solutions to these problems commercially available\cite{infleqtion, Japan_Cell, Prec_Glassblowing} by employing pure glass cells as the main scientific vacuum chamber. These chambers excel in the aforementioned parameters, while also contributing to minimizing the SWaP-budget (size, weight and power consumption), which is the main design driver in most modern quantum applications. While being the optimal solution for most experiments with cold atoms, the downsides to this vacuum technology are the limited availability, the rather high price tag, especially when custom geometries are involved, and technical challenges in achieving a double sided AR-coating in- and outside of the vacuum system. Other commercial material platforms, such as ceramic-based technologies\cite{Kyocera, Umicore, MDC}, are available in industrial UHV contexts but are in general not tailored towards flexible laboratory-scale realizations of customizable science cells. Furthermore, such technologies are typically focused either on the realization of ceramic-metal interfaces or on permanently sealed vacuum assemblies, and do not include the integration of optical viewports required for quantum applications.
Consequently, several recent academic approaches have explored alternative materials and fabrication strategies for compact UHV chambers. Notable examples include brazing sapphire or fused-silica windows into titanium frames\cite{SapphireToTitanium,Lee2022}, additively manufactured metal vacuum chambers\cite{COOPER2021101898}, and ceramic-based systems such as sintered alumina chambers with bonded windows\cite{Test2021}. While these approaches demonstrate excellent vacuum performance or increased geometric flexibility, they generally require specialized fabrication infrastructure and, in some cases, extensive post-processing to achieve UHV compatibility.

In contrast, the present work focuses on a low-cost, easily machinable glass-ceramic body with adhesive-based bonding techniques, enabling customizable UHV-compatible science cell manufacturing while minimizing fabrication complexity. As a material of choice we choose the glass ceramic MACOR \cite{corning}, which is relatively low-cost, easy to machine and lightweight. Due to its great machinability, MACOR can be processed using standard CNC equipment commonly employed for metals, which substantially reduces fabrication effort. Its intrinsic properties of being non-magnetic and electrically isolating reduces interactions with external electric and magnetic fields to a minimum.
To be able to realize a complete UHV-system we developed procedures to connect and seal the MACOR-chamber to a conventional metal-based flange, as well as, sealing UV fused silica windows to the chamber.
Suitablility of the technology is proven by building a UHV cell (science cell) out of MACOR and connecting it to an existing ultra-high vacuum system of an experiment to study cold Dysprosium atoms \cite{Muehlbauer2018}. Here the pressure of the system over a time of more than one year was kept constant, and cold atoms could be optically transported into the chamber.
Finally, we present an outlook how this technology can contribute to the growing endeavors towards more mobile, transportable and compact applications in quantum sensing and space applications \cite{Becker2018,Grotti2018,Frye2021,Qiao_2023,Deng2024} substituting MACOR with a different, heat-resistant ultra-stable glass ceramic, ZERODUR.  

\begin{figure}[h]
    \centering
    \includegraphics[width=0.4\textwidth]{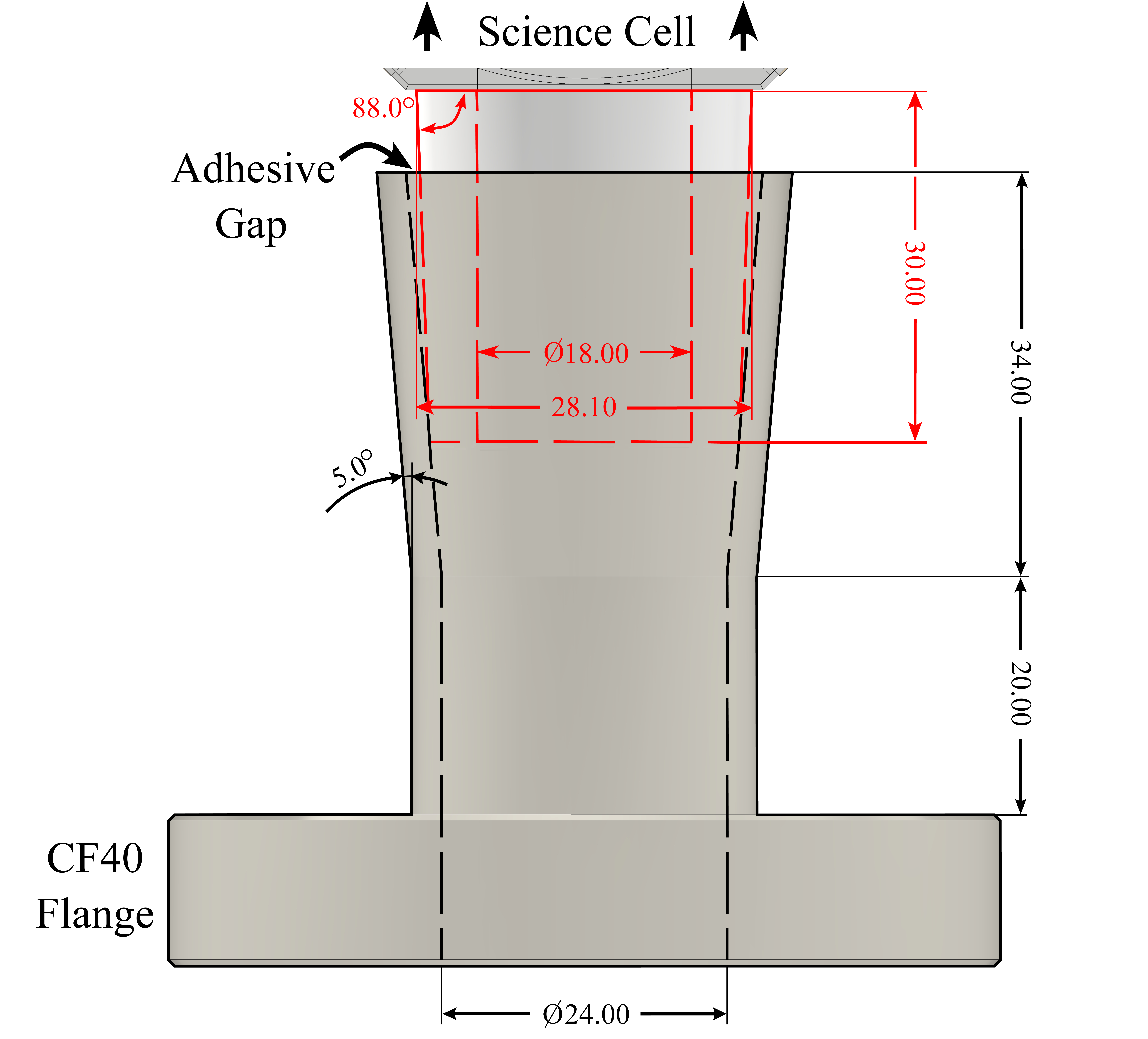}
    \caption{Technical drawing of the metal flange to MACOR chamber connection. The wall thickness of the metal part was chosen to be 2.5\,mm, while the glass ceramic conical extrusion has a minimum wall thickness of 4\,mm.}
    \label{fig:flangedesigntechnical}
\end{figure}

\begin{figure}[!tbp]
  \centering
  \subfloat[]{\label{fig:testkammer}%
    \includegraphics[width=0.45\linewidth]{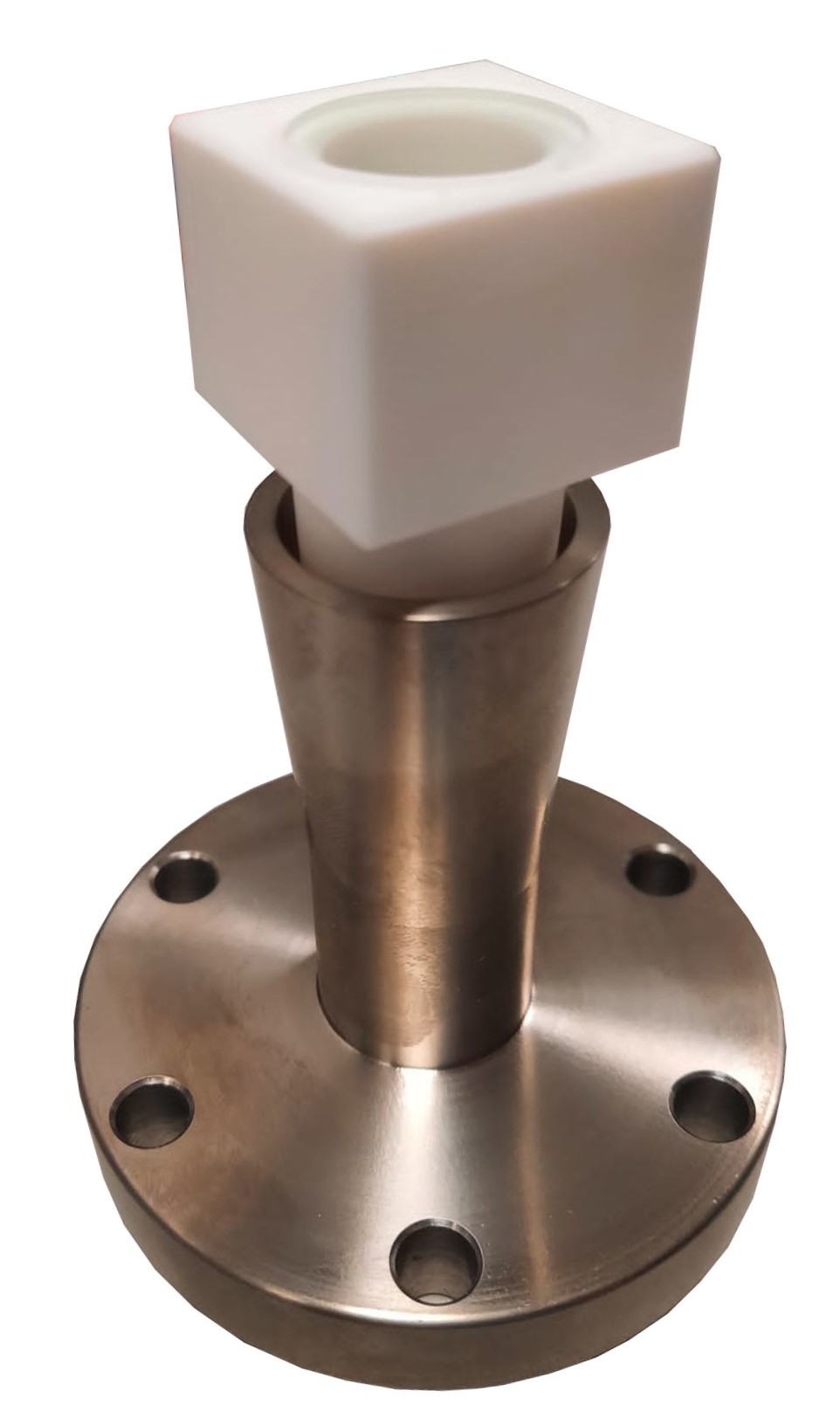}}
  \hfill
  \subfloat[]{\label{fig:testkammergeklebt}%
    \includegraphics[width=0.45\linewidth]{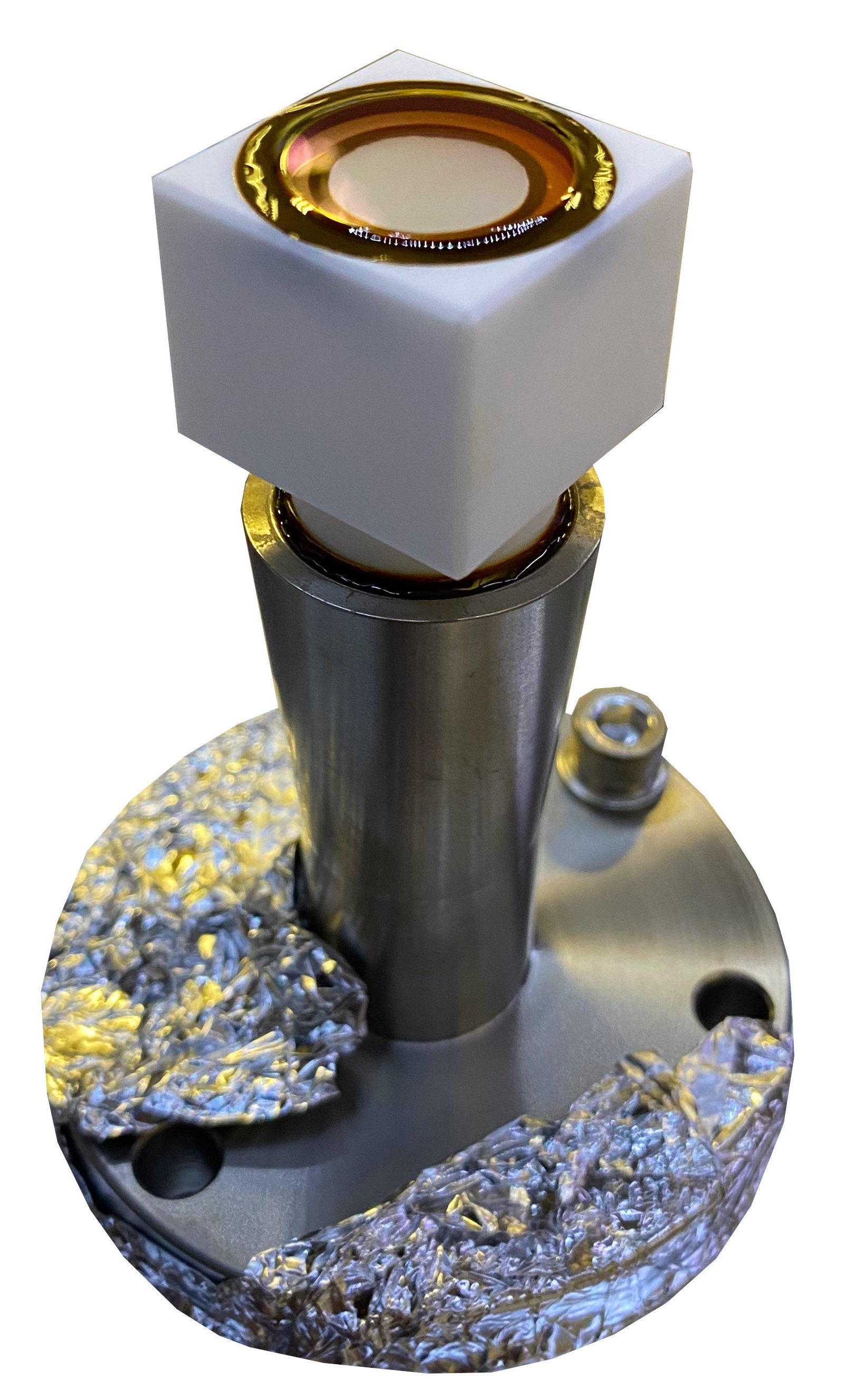}}
  \caption{(a) Photograph of the glass-ceramic to metal flange technology. The lower part shows the metallic CF40 flange adapter with a conical mounting tube. The test cell (25\,$\times$\,25\,$\times$\,20\,mm) made out of MACOR glass-ceramic can be easily inserted, after which an adhesive-based sealing of the conical mating, as well as the optical viewport, is established. (b) Photograph of the test chamber after the finished adhesive-based sealing procedure, here performed using Epotek~353ND.}
  \label{fig:flange-tech}
\end{figure}

To be able to connect commercial devices such as pumps, gauges or even metal based parts of an already existing vacuum system to the science chamber, a hermetic flange connection between the glass-ceramic chamber and metal based components is required. As glass ceramics are too brittle to apply sufficient force to deform a copper gasket to a degree where the connection becomes UHV tight, like in CF-type flanges, an alternative sealing method had to be found that strongly reduces or removes mechanical stress from the sealing connection. One of these approaches, most commonly found to seal brittle materials like optical glasses to metallic viewports, involves Indium sealing rings deformed in between two flat surfaces \cite{SAEKI1989563}.
This approach requires great surface cleanliness as well as flatness, the latter being a property hard to obtain using glass-ceramics, unless expensive polishing of the flange surfaces is performed. Additionally, a deep understanding of the maximum clamping forces in dependency of the material thicknesses and properties is required to avoid irreversible damage to the vacuum components. Alternative joining techniques, such as active brazing, offer high mechanical strength and bakeability, but typically require high processing temperatures, vacuum furnaces and careful material selection, which can limit compatibility with machinable glass ceramics and increase fabrication complexity.
We developed a conical flange design in which we can establish an adhesive based connection between a glass-ceramic and metal that is not reliant on the material surfaces touching under significant amounts of mechanical stress.
The schematic of the design is laid out in figure \ref{fig:flangedesigntechnical}. It consists of two conical mating parts with a slightly different opening angle on both sides of the to be connected parts. The difference leads to a gap between both materials, that can be filled with an adhesive to form a mechanically stable and hermetic sealing connection. A reducing angle of the glass-ceramic part was chosen to maximize the wall thickness around the adapter to chamber intersection, while also contributing to the SWaP-budget by continuously reducing the wall thickness over the length of the adapter. In the depicted case, the conical parts' overlap amounted to 2/3 compared to the respective conical parts' total length, which made the gap easily accessible with a 1\,mm diameter syringe for adhesive application. For the flange dimension we chose DNCF40, as this can be seen as the commercial standard for most medium sized vacuum systems, and it also suited our available vacuum equipment. In general, the presented dimensions can be arbitrarily scaled to suit the experimental requirements. The lengths shown in the schematic are the exact proportions used  for the construction of the science cell, while for the test chamber, shown in figure \ref{fig:testkammer}, the metallic flange extrusion and the chamber were scaled down to reduce material cost. The reduced clear aperture of the flange was chosen to match the optical dipole trap beam geometry. In general, however, the dimensions can be scaled to fully utilize the maximum CF40 opening if required.
For the flange materials, Titanium was selected as the metal part, while MACOR was the material of choice for the glass-ceramic counterpart. The difference in the coefficients of thermal expansion (CTE) of these two materials results in a maximum mismatch of only $2\cdot10^{-7}\,\text{K}^{-1}$ between 20\,$^\circ\mathrm{C}$ and 150\,$^\circ\mathrm{C}$ \cite{corning,Titanium}, reducing the amount of stress around the circular touching area of both materials in case of external temperature changes (fluctuations, baking, etc.). To test this connection concept a MACOR test chamber was constructed (see also figure \ref{fig:testkammer}) and inserted into a Titanium flange utilizing the conical extrusion from the cell body. The adhesive gap between flange and cell is clearly visible and a singular window is placed onto the recess of the chamber.

\begin{figure}[h]
    \centering
    \includegraphics[width=0.4\textwidth]{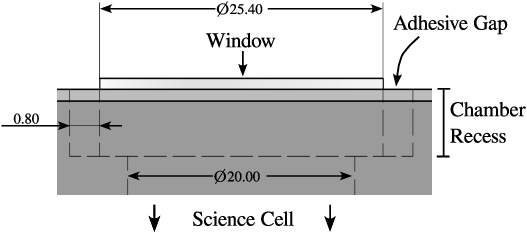}
    \caption{Technical drawing of the glass window to MACOR chamber connection for a 1" window. The diameter of the chamber recess and window are chosen such, that the clear aperture is fully covered and the depicted gap is formed. The thickness of the window (3.175\,mm) is slightly larger than the chamber recess depth (3.0\,mm) to prevent adhesive from spilling onto the window surface during the application process.}
    \label{fig:windowdesign}
\end{figure}

The aim of the presented glass-ceramic to metal sealing technology is to reduce mechanical stress in between these materials as much as possible, while maintaining high control and reproducibility in the adhesives' spread. With the same motivation the glass-ceramic to window sealing was developed, as depicted in figure \ref{fig:windowdesign}.
Here, a circular window is placed into a recess of the MACOR chamber body covering the clear aperture into the chamber. The diameter of the recess is chosen slightly larger than the window's diameter, resulting in the depicted gap forming around the window that can be filled with adhesive. This approach was preferred over placing a window onto a flat surface already covered in adhesive, as it again yields more reproducible results concerning adhesive spread, especially in geometries with multiple windows. The underlying concept with this method is the utilization of thermal expansion of the adhesive in the curing process, spreading into the window and glass-ceramics interface homogeneously, removing trapped gas on the way. For the window material UV fused silica was selected as its comparable low CTE reduces mechanical stress in applications where high power lasers are used.

The foremost criteria for the selection of adhesives was to minimize the outgassing rate, measured by the total mass loss (TML) after being exposed to 200\,°C for 24 hours in vacuum. This ensures long-term stability and longevity of the vacuum sealing. Moreover, a suitable viscosity range had to be found, which allows the adhesive to spread between the chamber to window interface, but prevents leaking of excessive amounts of adhesive into the chamber, which could lead to the obstruction of the optical aperture due to surface cohesion between adhesive and window. The most promising candidates selected for further testing were 353ND, 377 and H77S, which are well-established UHV adhesives in industry and technology, distributed by Epoxy Technology with their relevant properties displayed in table \ref{tableprops}.

\begin{figure}[h]
    \centering
    \includegraphics[width=0.4\textwidth]{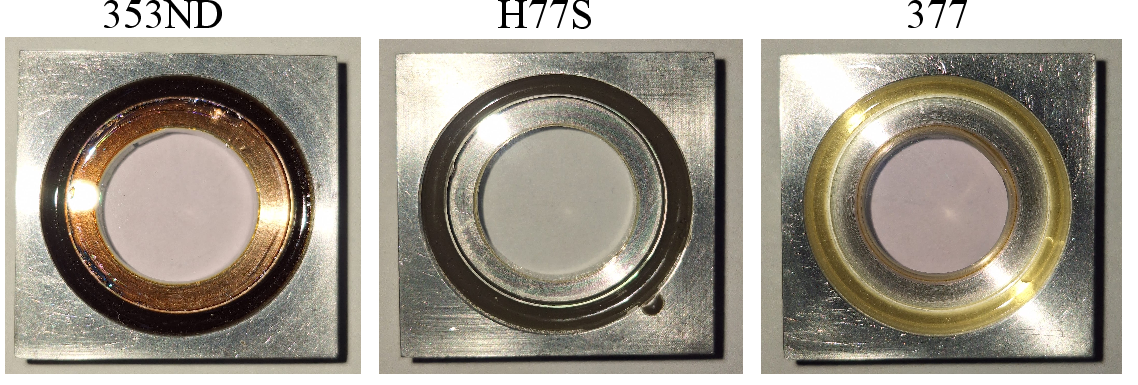}
    \caption{Photographs of the aluminium viewports including a window after curing of the stated adhesives. 377 shows an increased leakage of adhesive into the viewport, indicated by the ring formation around the clear aperture.}
    \label{fig:adhesiveleak}
\end{figure}

To test the aforementioned adhesive spreading properties, several windows were placed onto aluminium optical viewports with unpolished surfaces. This allows for microscopic slits to form at the contact area of window and aluminum surface, the same as would be expected when substituting the aluminum with a porous glass-ceramic. During the adhesive curing process, thermal expansion pushes the adhesive into these slits, slowly diffusing towards the viewport's aperture. The three adhesives were prepared in accordance with their respective datasheets \cite{353datasheet,H77datasheet,377datasheet}, and then applied into the adhesive gap with a syringe and cured. The result of this process for the three test adhesives is depicted in figure \ref{fig:adhesiveleak}. The significantly lower viscoscity of 377 allowed a substantial amount of adhesive to leak into the window-aluminium interface and accumulate around the clear aperture, potentially reducing the optical access in experimental applications. By this 377 was ruled out as a potential candidate for our application.
For the metal to glass-ceramic sealing no influence of potential adhesive leaks on the vacuum or optical performance is expected due to the chosen geometry. In subsequent test chamber sealings (see also figure \ref{fig:testkammergeklebt})  no leakages around the circular touching area of metal and glass-ceramic could be observed.

\begin{table}[!t]
\footnotesize
\renewcommand{\arraystretch}{1.2}
\caption{Overview of important properties of the adhesives to be tested \cite{353datasheet,H77datasheet,377datasheet}.}
\label{tableprops}
\centering
\begin{ruledtabular}
\begin{tabular*}{\columnwidth}{@{\extracolsep{\fill}} lccc}
Adhesive                 & ET 353ND     & ET H77S       & ET 377      \\
\hline
Viscosity [cP]           & 3,000--5,000 & 950--1,500    & 150--300    \\
Weight loss @ 200\,$^\circ\mathrm{C}$ [\%] & 0.22         & \textless{}0.05 & 0.06        \\
\end{tabular*}
\end{ruledtabular}
\end{table}

With the Epotek 353ND and H77S confirmed as potential candidates for the UHV sealing procedure, we constructed MACOR test chambers featuring one optical viewport and a flange connection, depicted in figure \ref{fig:testkammer}. The conical metallic flange was CNC-machined out of a solid Titanium piece, with all dimensions scaled by a constant factor to maintain the relative proportions depicted in figure \ref{fig:flangedesigntechnical}. 
Prior to application, the adhesives were first mixed according to their datasheets and then transitioned into test tubes and degassed using an ultrasonic bath for 5 minutes. Special care was taken to not degas for extended times, since it was found to initiate the curing process after around 20 minutes in the ultrasonic bath. 
Following the degassing, the adhesives were filled into syringes and applied to the respective adhesive gaps of metal to glass-ceramic and glass-ceramic to window connections. The sealings were precured at room temperature for two days to start the gelling of the adhesive. Afterwards, the test chambers were placed into an oven and cured at 80\,$^\circ\mathrm{C}$, with a heating rate of $0.25\, \frac{^\circ\mathrm{C}}{\text{min}}$, followed by 8\,h at 80\,$^\circ\mathrm{C}$, and a subsequent cooling with the same rate. This approach is recommended by the manufacturer for curing at lower temperatures. The deviation from the faster curing process of 1\,h at 150\,$^\circ\mathrm{C}$ was chosen, due to the windows being more prone to break after the fast curing process, when external force was applied. This indicates that more mechanical stress was induced during the heating and subsequent cooling process.

Using this procedure, two test chambers were assembled (see also figure \ref{fig:testkammergeklebt} for the case of 353ND), employing Epotek 353ND and H77S adhesives as sealants, respectively. As a preliminary test for UHV suitability, both chambers underwent a helium leak test. To that end an Agilent G8610B Helium leakage seeker with a minimal detectable leakage rate of $Q_L = 10^{-11} \text{mbar} \cdot  \frac{\text{l}}{\text{s}}$ was used.
Both chambers were attached to the Helium leakage seeker via a standard CF40 flange connection using Copper sealing gaskets. After generously applying Helium onto the test chambers with a pressure nozzle for extended periods, we could not detect any leakages, resulting in a lower limit leakage rate of $Q_L < 10^{-11} \text{mbar} \cdot  \frac{\text{l}}{\text{s}}$.
After qualifying both chambers as sufficiently hermetic for UHV applications, they were successively integrated into a vacuum system, consisting of an $300\frac{\text{l}}{\text{s}}$ turbomolecular pump and a $10\frac{\text{l}}{\text{s}}$ ion getter pump. In table \ref{tablepressure}, the final pressures are noted, with all measurements taken using a Pfeiffer IKR 270 vacuum gauge after roughly 14 days bake-out at 60\,$^\circ\mathrm{C}$. The bake-out temperature was kept below the adhesive curing temperature to avoid inducing additional mechanical stress. We found that both chamber sealing procedures produced a UHV suitable sealing of the chambers, with similiar base pressures as a previous reference measurement, where the chamber was replaced with a standard CF40 blind flange. It should be noted, that the blind flange reference measurement was taken after a higher bakeout temperature of 200\,$^\circ\mathrm{C}$ to clean the vacuum system of undesirable gas loads before investigating the adhesive sealings.
Though the previous tests qualified both adhesives for a further use, we ultimately selected Epotek 353ND as the adhesive of choice for the construction of the glass-ceramic science cell due to its greater availability at the time of testing.

\begin{table}[!t]
\renewcommand{\arraystretch}{1.2}
\centering
\caption{Base pressures achieved for different test chambers, as well as a reference blind flange measurement. All measurements were taken after around 14 days of pumping time. The blind flange configuration was baked out at 200\,$^\circ\mathrm{C}$ to initially clean the vacuum system of undesired gas loads. While the adhesive-based vacuum chambers were only baked out at 60\,$^\circ\mathrm{C}$.}
\label{tablepressure}
\begin{ruledtabular}
\begin{tabular*}{8.5cm}{@{\extracolsep{\fill}} l r}
\multicolumn{2}{l}{Base vacuum pressure after bakeout} \\
\hline
Blind flange            & $(3.0 \pm 1.0)\times 10^{-10}\,\mathrm{mbar}$ \\
Test chamber ET 353ND   & $(4.3 \pm 1.3)\times 10^{-10}\,\mathrm{mbar}$ \\
Test chamber ET H77S    & $(4.7 \pm 1.4)\times 10^{-10}\,\mathrm{mbar}$ \\
Science cell ET 353ND   & $(4.1 \pm 1.4)\times 10^{-10}\,\mathrm{mbar}$ \\
\end{tabular*}
\end{ruledtabular}
\end{table}

\begin{figure}[h]
    \centering
    \includegraphics[width=0.4\textwidth]{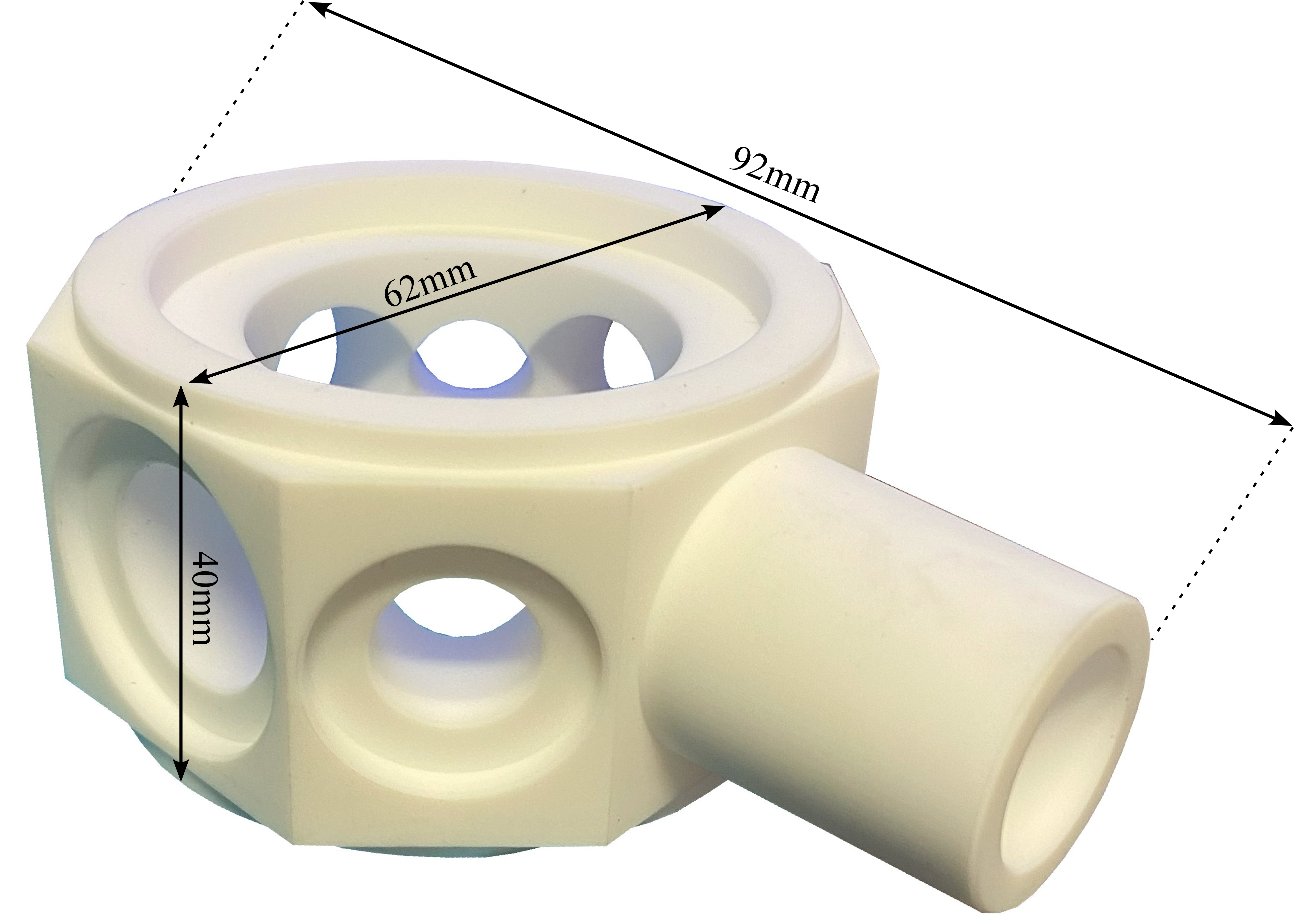}
    \caption{Picture of the MACOR science cell. Its full dimensions are denoted in the figure, featuring a 62x62x40\,mm octagonal cell body and a 30\,mm conical flange adapter. The nine optical window recesses are designed to hold commercial 0.75", 1" and 2" windows.}
    \label{fig:sciencecellausgestellt2}
\end{figure}

\begin{figure}[h]
    \centering
    \includegraphics[width=0.5\textwidth]{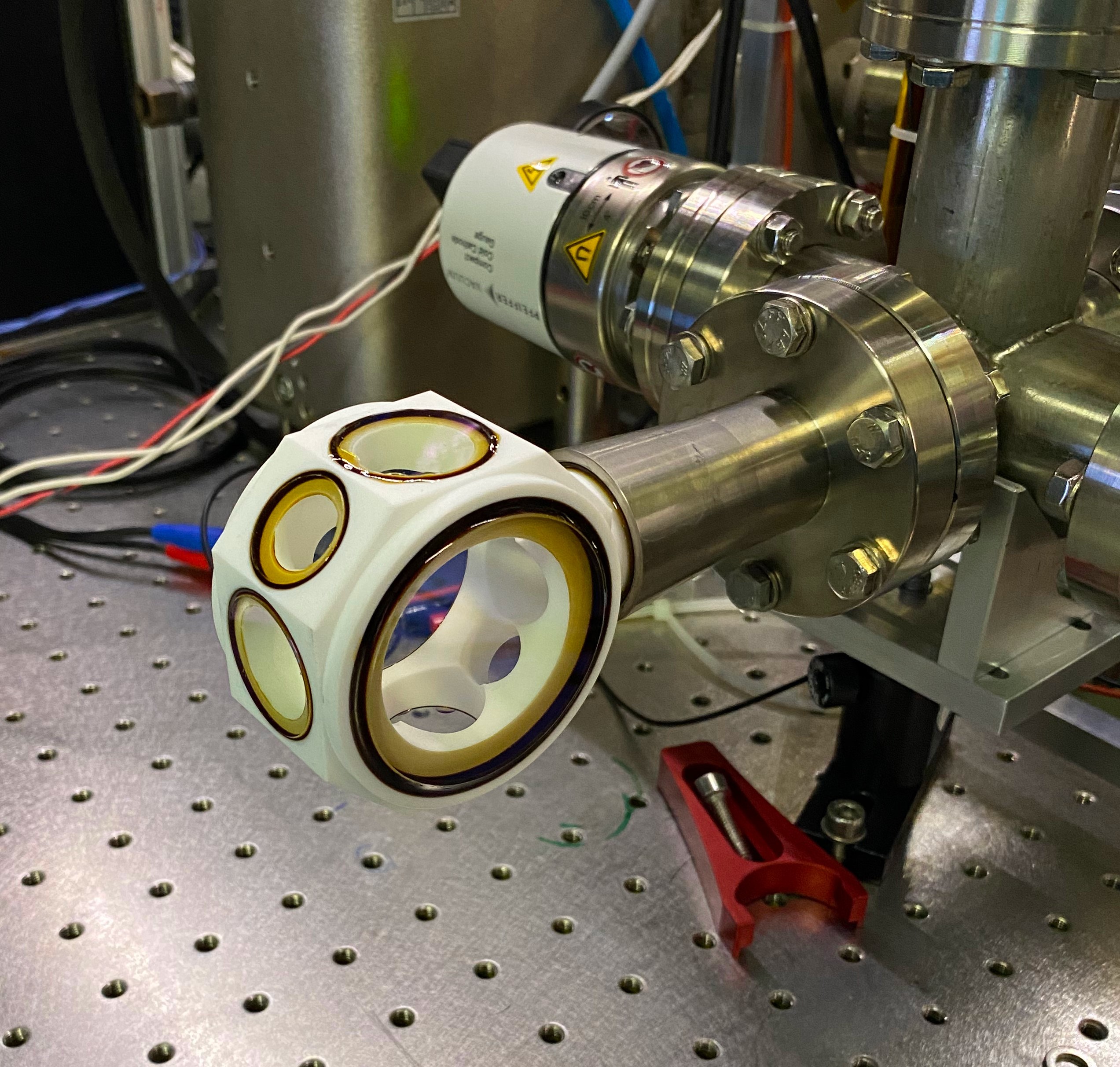}
    \caption{Photograph of the MACOR science cell mounted in the experimental system. After an initial bake-out at 60\,$^\circ\mathrm{C}$ for two weeks, the lower limit of the vacuum sensor of $\leq 1\cdot10^{-10}\,\text{mbar}$ was reached.}
    \label{fig:sciencecellinsystem}
\end{figure}

To demonstrate scalability and suitability for arbitrary experimental platforms, we designed and constructed a MACOR science cell depicted in figure \ref{fig:sciencecellausgestellt2}, featuring nine optical viewports of various sizes, which will be implemented into a cold atom experiment operating with Dysprosium atoms \cite{DyExperiment,PhysRevResearch.6.023147}. 
The cell features two 40.0\,mm viewports for high numerical aperture applications, as well as three 20.0\,mm and four 13.6\,mm viewports for atom transport, absorption/fluorescence imaging and state preparation.
After cleaning the chamber body in an ultrasonic bath, all windows were glued to the science cell successively according to the aforementioned procedure, after which the flange connection sealing was established.
Subsequently, the chamber was cleaned by flushing it with isopropyl alcohol. Special care was taken to not leave any large drops of the cleaning liquid, as well as not to use harsh chemicals such as acetone. The cleaned science cell was then mounted onto the Helium leakage seeker, which again showed a lower limit leakage rate of $Q_L < 10^{-11} \text{mbar} \cdot  \frac{\text{l}}{\text{s}}$. After the verification of its Helium hermeticity, the science cell was first attached to the test vacuum system, with which a base pressure of $4.1 \pm 1.4 \cdot 10^{-10}\,\text{mbar}$ was observed. This agrees very well with the prior test chamber measurements, qualifying the science cell for the final integration into the Dysprosium experiment's vacuum system. Figure \ref{fig:sciencecellinsystem} shows the cell after integration connected to a stainless steel DNCF40 5-way cross. The 5-way cross allows for the connection of a UHV-valve to separate the science cell system from the main chamber and an angle valve for separate pre-pumping before the cell is baked at 60\,$^\circ\mathrm{C}$ for two weeks. Additionally, a SAES Getters NEXTorr Z100 combination pump was connected to the system increasing the pumping speed to 150\,l/s (H2). After getter activation and bake-out a final pressure of $8.9 \pm 1.0 \cdot 10^{-11}$ mbar could be measured with a Pfeiffer IKR 270 vacuum gauge attached to the 5-way cross. This gauge had to be removed later on, due to its strong magnetic field influencing the highly magnetic Dysprosium atoms. Nevertheless, the ion pump controller of the NEXTorr Z100 consistently displayed a pressure of $\leq 1\cdot10^{-10}\,\text{mbar}$ measured via the ion pump supply current, which has not changed as of publishing this report (over a time period of 23 months).
In this time, Dysprosium atoms could be transported from the MOT chamber into the science cell over a distance of 41\,cm utilizing a focus shifted optical dipole trap with transport efficiencies exceeding 60\%. The transport mechanism is detailed in \cite{PhysRevResearch.6.023147}.
As of today, no degradation of experimental performance indicative of increased background scattering collisions has been observed following the installation of the science cell, further indicating similar vacuum performance to established steel chambers.
This emphasizes the maturity of the technology as an alternative to conventional science cell systems.

In summary, we presented a new approach for ultra-high vacuum sealing of CF-based metal flanges, as well as optical viewports to a custom-geometry MACOR glass-ceramic science cell utilizing adhesive-based bonding. Competitive leakage rates and base pressures, when compared to conventional vacuum technology, were achieved. The technology enables the reliable construction of glass-ceramic vacuum chambers for various applications, such as magneto-optical trapping or optical dipole trapping of atomic ensembles. The robustness of the sealing method was demonstrated by integration of a multiple viewport custom-geometry scientific chamber into an existing cold atom experiment and optical transport of atoms into it. With a stable base pressure maintained for more than a year, the longevity of the technology was demonstrated, indicating its suitability as a reliable long-term solution for scientific vacuum chambers. Additionally, the possibility of using commercial optical windows as viewports, offers great flexibility for customized window coatings, enhancing the range of experimental applications. The possibility of arbitrarily selecting cell geometry, window coating and sizes, as well as flange dimensions serves as a proof of the high grade of flexibility in the presented technology, making it a serious alternative to established science cell technology.

The technology and techniques demonstrated in this paper are not limited to the materials of the presented science cell. They are also expected to be compatible with various ceramic, glass or glass-ceramic materials, such as Zerodur or Alumina (Al\textsubscript{2}O\textsubscript{3}), potentially broadening the range of materials suitable for specialized applications. Consequently, the next step is to apply the presented technology to a vacuum chamber, where both the chamber body and optical windows are made of ZERODUR, a glass-ceramic with low helium permeability and a negligible CTE\cite{zerodur_schott}, thereby providing resilience against temperature gradients over large ranges, a property advantageous for quantum sensing in space applications\cite{Frye2021,2022SPIE12180E..5NK,Duncker:14}.

\vspace{\baselineskip}
This work is supported by the German Space Agency DLR with funds provided by the Federal Ministry for economic affairs and energy (BMWE) under grant numbers DLR 50WP1433, 50WP2103 and 50WM2266B. We acknowledge financial support by the
Deutsche Forschungsgemeinschaft (DFG, German Research Foundation) Project-ID 429529648--TRR 306 QuCoLiMa ("Quantum Cooperativity of Light and Matter") and the Quanten-Initiative Rheinland-Pfalz (QUIP, Quantum-Initiative Rhineland-Palatinate).

\section*{AUTHOR DECLARATIONS}
\subsection*{Conflict of Interest}

The authors have no conflicts to disclose.

\subsection*{Author Contributions}
M.P. selected the materials, prepared the design of test chamber/science cell and contributed to the vacuum tests, science cell implementation and paper preparation; S.B. conceptualized the flange and window sealing technology, led the vacuum cell assembly and testing, and supplied the initial paper draft; D.L.B. contributed to the vacuum tests and revising the paper; I.V. and R.S. contributed to the experimental implementation and testing of the science cell; O.H and K.S acquired funding, supported development and supervised design concepts in the early experimental phase; A.W. and P.W. acquired funding, supervised the experimental work and revised the manuscript. \\

\section*{Data Availability Statement}

The data that support the findings of this study are available from the corresponding author upon reasonable request.

\section*{REFERENCES}
\nocite{*}
\bibliography{aipsamp}

@PREAMBLE{
 "\providecommand{\noopsort}[1]{}" 
 # "\providecommand{\singleletter}[1]{#1}%" 
}

@Article{Schleich2016,
author={Schleich, Wolfgang P.
and Ranade, Kedar S.
and Anton, Christian
and Arndt, Markus
and Aspelmeyer, Markus
and Bayer, Manfred
and Berg, Gunnar
and Calarco, Tommaso
and Fuchs, Harald
and Giacobino, Elisabeth
and Grassl, Markus
and H{\"a}nggi, Peter
and Heckl, Wolfgang M.
and Hertel, Ingolf-Volker
and Huelga, Susana
and Jelezko, Fedor
and Keimer, Bernhard
and Kotthaus, J{\"o}rg P.
and Leuchs, Gerd
and L{\"u}tkenhaus, Norbert
and Maurer, Ueli
and Pfau, Tilman
and Plenio, Martin B.
and Rasel, Ernst Maria
and Renn, Ortwin
and Silberhorn, Christine
and Schiedmayer, J{\"o}rg
and Schmitt-Landsiedel, Doris
and Sch{\"o}nhammer, Kurt
and Ustinov, Alexey
and Walther, Philip
and Weinfurter, Harald
and Welzl, Emo
and Wiesendanger, Roland
and Wolf, Stefan
and Zeilinger, Anton
and Zoller, Peter},
title={Quantum technology: from research to application},
journal={Applied Physics B},
year={2016},
month={Apr},
day={27},
volume={122},
number={5},
pages={130},
issn={1432-0649},
doi={10.1007/s00340-016-6353-8},
url={https://doi.org/10.1007/s00340-016-6353-8}
}

@Article{Kaltenbaek2021,
author={Kaltenbaek, Rainer
and Acin, Antonio
and Bacsardi, Laszlo
and Bianco, Paolo
and Bouyer, Philippe
and Diamanti, Eleni
and Marquardt, Christoph
and Omar, Yasser
and Pruneri, Valerio
and Rasel, Ernst
and Sang, Bernhard
and Seidel, Stephan
and Ulbricht, Hendrik
and Ursin, Rupert
and Villoresi, Paolo
and van den Bossche, Mathias
and von Klitzing, Wolf
and Zbinden, Hugo
and Paternostro, Mauro
and Bassi, Angelo},
title={Quantum technologies in space},
journal={Experimental Astronomy},
year={2021},
month={Jun},
day={01},
volume={51},
number={3},
pages={1677-1694},
issn={1572-9508},
doi={10.1007/s10686-021-09731-x},
url={https://doi.org/10.1007/s10686-021-09731-x}
}

@article{RevModPhys.94.041001,
  title = {Colloquium: Atomtronic circuits: From many-body physics to quantum technologies},
  author = {Amico, Luigi and Anderson, Dana and Boshier, Malcolm and Brantut, Jean-Philippe and Kwek, Leong-Chuan and Minguzzi, Anna and von Klitzing, Wolf},
  journal = {Rev. Mod. Phys.},
  volume = {94},
  issue = {4},
  pages = {041001},
  numpages = {30},
  year = {2022},
  month = {Nov},
  publisher = {American Physical Society},
  doi = {10.1103/RevModPhys.94.041001},
  url = {https://link.aps.org/doi/10.1103/RevModPhys.94.041001}
}

@Article{O'Brien2009,
author={O'Brien, Jeremy L.
and Furusawa, Akira
and Vu{\v{c}}kovi{\'{c}}, Jelena},
title={Photonic quantum technologies},
journal={Nature Photonics},
year={2009},
month={Dec},
day={01},
volume={3},
number={12},
pages={687-695},
abstract={We have just witnessed the birth of the first quantum technology based on encoding information in light for quantum key distribution. The quantum nature of light seems destined to continue to have a central role in future technologies. Here we provide a broad review of photonics for quantum technologies touching on topics including secure communication with photons, quantum information processing, quantum lithography and integrated quantum photonics.},
issn={1749-4893},
doi={10.1038/nphoton.2009.229},
url={https://doi.org/10.1038/nphoton.2009.229}
}

@article{
doi:10.1073/pnas.1419326112,
author = {Gershon Kurizki  and Patrice Bertet  and Yuimaru Kubo  and Klaus Mølmer  and David Petrosyan  and Peter Rabl  and Jörg Schmiedmayer },
title = {Quantum technologies with hybrid systems},
journal = {Proceedings of the National Academy of Sciences},
volume = {112},
number = {13},
pages = {3866-3873},
year = {2015},
doi = {10.1073/pnas.1419326112},
URL = {https://www.pnas.org/doi/abs/10.1073/pnas.1419326112},
eprint = {https://www.pnas.org/doi/pdf/10.1073/pnas.1419326112}}

@Article{Pesce2020,
author={Pesce, Giuseppe
and Jones, Philip H.
and Marag{\`o}, Onofrio M.
and Volpe, Giovanni},
title={Optical tweezers: theory and practice},
journal={The European Physical Journal Plus},
year={2020},
month={Dec},
day={07},
volume={135},
number={12},
pages={949},
issn={2190-5444},
doi={10.1140/epjp/s13360-020-00843-5},
url={https://doi.org/10.1140/epjp/s13360-020-00843-5}
}

@misc{infleqtion,
  author = {{Infleqtion, Inc.}},
  year   = {2025},
  note   = {3030 Sterling Circle, Boulder, CO 80301, USA; \url{https://www.infleqtion.com/}}
}

@misc{Japan_Cell,
  author = {{Japan Cell Co., Ltd.}},
  year   = {2025},
  note   = {Machida Technopark, 2-2-5-11, Oyamagaoka, Machida-shi, Tokyo 194-0215, Japan; \url{https://www.jpcell.co.jp/en/}}
}

@misc{Prec_Glassblowing,
  author = {{Precision Glassblowing}},
  year   = {2025},
  note   = {14775 E. Hinsdale Ave., Centennial, CO 80112, USA; \url{https://precisionglassblowing.com/}}
}

@misc{Kyocera,
  author = {{Kyocera}},
  year   = {2026},
  note   = {6 Takeda Tobadono-cho, Fushmi-ku, Kyoto, Japan 612-8501; \url{https://global.kyocera.com/}}
}

@misc{Umicore,
  author = {{Umicore}},
  year   = {2026},
  note   = {Broekstraat 31, 1000 Brussel, Belgium; \url{https://www.umicore-ceramics.com/}}
}

@misc{MDC,
  author = {{MDC Precision}},
  year   = {2026},
  note   = {30962 Satana Street, Hayward, California 94544, USA; \url{https://www.mdcprecision.com/}}
}

@misc{corning,
author = {{Corning Inc.}},
year   = {2025},
title = {MACOR® Machinable Glass Ceramic For Industrial Applications},
note = {One Riverfront Plaza, New York, 14831, USA; \url{https://www.corning.com/}},
}

@Article{Muehlbauer2018,
author={M{\"u}hlbauer, Florian
and Petersen, Niels
and Baumg{\"a}rtner, Carina
and Maske, Lena
and Windpassinger, Patrick},
title={Systematic optimization of laser cooling of dysprosium},
journal={Applied Physics B},
year={2018},
month={May},
day={29},
volume={124},
number={6},
pages={120},
issn={1432-0649},
doi={10.1007/s00340-018-6981-2},
url={https://doi.org/10.1007/s00340-018-6981-2}
}

@Article{Becker2018,
author={Becker, Dennis
and Lachmann, Maike D.
and Seidel, Stephan T.
and Ahlers, Holger
and Dinkelaker, Aline N.
and Grosse, Jens
and Hellmig, Ortwin
and M{\"u}ntinga, Hauke
and Schkolnik, Vladimir
and Wendrich, Thijs
and Wenzlawski, Andr{\'e}
and Weps, Benjamin
and Corgier, Robin
and Franz, Tobias
and Gaaloul, Naceur
and Herr, Waldemar
and L{\"u}dtke, Daniel
and Popp, Manuel
and Amri, Sirine
and Duncker, Hannes
and Erbe, Maik
and Kohfeldt, Anja
and Kubelka-Lange, Andr{\'e}
and Braxmaier, Claus
and Charron, Eric
and Ertmer, Wolfgang
and Krutzik, Markus
and L{\"a}mmerzahl, Claus
and Peters, Achim
and Schleich, Wolfgang P.
and Sengstock, Klaus
and Walser, Reinhold
and Wicht, Andreas
and Windpassinger, Patrick
and Rasel, Ernst M.},
title={Space-borne Bose--Einstein condensation for precision interferometry},
journal={Nature},
year={2018},
month={Oct},
day={01},
volume={562},
number={7727},
pages={391-395},
issn={1476-4687},
doi={10.1038/s41586-018-0605-1},
url={https://doi.org/10.1038/s41586-018-0605-1}
}

@Article{Grotti2018,
author={Grotti, Jacopo
and Koller, Silvio
and Vogt, Stefan
and H{\"a}fner, Sebastian
and Sterr, Uwe
and Lisdat, Christian
and Denker, Heiner
and Voigt, Christian
and Timmen, Ludger
and Rolland, Antoine
and Baynes, Fred N.
and Margolis, Helen S.
and Zampaolo, Michel
and Thoumany, Pierre
and Pizzocaro, Marco
and Rauf, Benjamin
and Bregolin, Filippo
and Tampellini, Anna
and Barbieri, Piero
and Zucco, Massimo
and Costanzo, Giovanni A.
and Clivati, Cecilia
and Levi, Filippo
and Calonico, Davide},
title={Geodesy and metrology with a transportable optical clock},
journal={Nature Physics},
year={2018},
month={May},
day={01},
volume={14},
number={5},
pages={437-441},
issn={1745-2481},
doi={10.1038/s41567-017-0042-3},
url={https://doi.org/10.1038/s41567-017-0042-3}
}

@Article{Frye2021,
author={Frye, Kai
and Abend, Sven
and Bartosch, Wolfgang
and Bawamia, Ahmad
and Becker, Dennis
and Blume, Holger
and Braxmaier, Claus
and Chiow, Sheng-Wey
and Efremov, Maxim A.
and Ertmer, Wolfgang
and Fierlinger, Peter
and Franz, Tobias
and Gaaloul, Naceur
and Grosse, Jens
and Grzeschik, Christoph
and Hellmig, Ortwin
and Henderson, Victoria A.
and Herr, Waldemar
and Israelsson, Ulf
and Kohel, James
and Krutzik, Markus
and K{\"u}rbis, Christian
and L{\"a}mmerzahl, Claus
and List, Meike
and L{\"u}dtke, Daniel
and Lundblad, Nathan
and Marburger, J. Pierre
and Meister, Matthias
and Mihm, Moritz
and M{\"u}ller, Holger
and M{\"u}ntinga, Hauke
and Nepal, Ayush M.
and Oberschulte, Tim
and Papakonstantinou, Alexandros
and Perovs̆ek, Jaka
and Peters, Achim
and Prat, Arnau
and Rasel, Ernst M.
and Roura, Albert
and Sbroscia, Matteo
and Schleich, Wolfgang P.
and Schubert, Christian
and Seidel, Stephan T.
and Sommer, Jan
and Spindeldreier, Christian
and Stamper-Kurn, Dan
and Stuhl, Benjamin K.
and Warner, Marvin
and Wendrich, Thijs
and Wenzlawski, Andr{\'e}
and Wicht, Andreas
and Windpassinger, Patrick
and Yu, Nan
and W{\"o}rner, Lisa},
title={The Bose-Einstein Condensate and Cold Atom Laboratory},
journal={EPJ Quantum Technology},
year={2021},
month={Jan},
day={04},
volume={8},
number={1},
pages={1},
issn={2196-0763},
doi={10.1140/epjqt/s40507-020-00090-8},
url={https://doi.org/10.1140/epjqt/s40507-020-00090-8}
}

@article{Qiao_2023,
doi = {10.1088/1742-6596/2651/1/012160},
url = {https://dx.doi.org/10.1088/1742-6596/2651/1/012160},
year = {2023},
month = {dec},
publisher = {IOP Publishing},
volume = {2651},
number = {1},
pages = {012160},
author = {Qiao, Zhongkun and Yuan, Peng and Zhou, Yin and Wu, Bin and Lin, Qiang},
title = {Marine Absolute Gravimetric Survey Based on Atomic Gravimeter},
journal = {Journal of Physics: Conference Series}
}

@article{Deng2024,
doi = {10.1088/1674-1056/ad4bc1},
url = {https://dx.doi.org/10.1088/1674-1056/ad4bc1},
year = {2024},
month = {jul},
publisher = {Chinese Physical Society and IOP Publishing Ltd},
volume = {33},
number = {7},
pages = {070602},
author = {Deng, Siminda and Ren, Wei and Xiang, Jingfeng and Zhao, Jianbo and Li, Lin and Zhang, Di and Wan, Jinyin and Meng, Yanling and Jiang, Xiaojun and Li, Tang and Liu, Liang and Lü, Desheng},
title = {Physics package based on intracavity laser cooling 87Rb atoms for space cold atom microwave clock},
journal = {Chinese Physics B}
}

@article{SAEKI1989563,
title = {Optical window sealed with indium for ultrahigh vacuum},
journal = {Vacuum},
volume = {39},
number = {6},
pages = {563-564},
year = {1989},
issn = {0042-207X},
doi = {https://doi.org/10.1016/0042-207X(89)90634-9},
url = {https://www.sciencedirect.com/science/article/pii/0042207X89906349},
author = {H Saeki and J Ikeda and H Ishimaru}
}

@article{Titanium,
    author = {P. Hidnert},
    title = {Thermal Expansion of Titanium},
    journal = {Journal of Research of the National Bureau of Standards},
    volume = {30},
    pages = {101-105},
    year = {1943},
    month = {02},
    doi = {10.6028/jres.030.008},
}

@misc{353datasheet,
title = {{EPO-TEK® 353ND, Technical Data Sheet}},
author = {{Epoxy Technology}},
month = {March},
year = {2023}
}

@misc{377datasheet,
title = {{EPO-TEK® 377, Technical Data Sheet}},
author = {{Epoxy Technology}},
month = {March},
year = {2023}
}

@misc{H77datasheet,
title = {{EPO-TEK® H77S, Technical Data Sheet}},
author = {{Epoxy Technology}},
month = {March},
year = {2023}
}

@article{DyExperiment,
  title = {Cooperative effects in dense cold atomic gases including magnetic dipole interactions},
  author = {Ba\ss{}ler, N. S. and Varma, I. and Proske, M. and Windpassinger, P. and Schmidt, K. P. and Genes, C.},
  journal = {Phys. Rev. Res.},
  volume = {6},
  issue = {2},
  pages = {023147},
  numpages = {12},
  year = {2024},
  month = {May},
  publisher = {American Physical Society},
  doi = {10.1103/PhysRevResearch.6.023147},
  url = {https://link.aps.org/doi/10.1103/PhysRevResearch.6.023147}
}

@article{PhysRevResearch.6.023147,
  title = {Cooperative effects in dense cold atomic gases including magnetic dipole interactions},
  author = {Ba\ss{}ler, N. S. and Varma, I. and Proske, M. and Windpassinger, P. and Schmidt, K. P. and Genes, C.},
  journal = {Phys. Rev. Res.},
  volume = {6},
  issue = {2},
  pages = {023147},
  numpages = {12},
  year = {2024},
  month = {May},
  publisher = {American Physical Society},
  doi = {10.1103/PhysRevResearch.6.023147},
  url = {https://link.aps.org/doi/10.1103/PhysRevResearch.6.023147}
}

@misc{zerodur_schott,
  author       = {{SCHOTT AG}},
  title        = {ZERODUR® Glass Ceramic – Technical Data Sheet},
  howpublished = {\url{https://www.schott.com/en-us/products/zerodur-p1000269/technical-details}},
  note         = {Hattenbergstrasse 10, Mainz, 55122, Germany},
  year         = {2025}
}

@INPROCEEDINGS{2022SPIE12180E..5NK,
       author = {{Krieg}, Janina and {Carr{\'e}}, Antoine and {D{\"o}hring}, Thorsten and {Hartmann}, Peter and {Hull}, Tony and {Jedamzik}, Ralf and {Westerhoff}, Thomas},
        title = "{The past decade of ZERODUR glass-ceramics in space applications}",
    booktitle = {Space Telescopes and Instrumentation 2022: Optical, Infrared, and Millimeter Wave},
         year = 2022,
       editor = {{Coyle}, Laura E. and {Matsuura}, Shuji and {Perrin}, Marshall D.},
       series = {Society of Photo-Optical Instrumentation Engineers (SPIE) Conference Series},
       volume = {12180},
        month = aug,
          eid = {121805N},
        pages = {121805N},
          doi = {10.1117/12.2628956},
       adsurl = {https://ui.adsabs.harvard.edu/abs/2022SPIE12180E..5NK}
}

@article{Duncker:14,
author = {Hannes Duncker and Ortwin Hellmig and Andr\'{e} Wenzlawski and Alexander Grote and Amir Jones Rafipoor and Mona Rafipoor and Klaus Sengstock and Patrick Windpassinger},
journal = {Appl. Opt.},
number = {20},
pages = {4468--4474},
publisher = {Optica Publishing Group},
title = {Ultrastable, Zerodur-based optical benches for quantum gas experiments},
volume = {53},
month = {Jul},
year = {2014},
url = {https://opg.optica.org/ao/abstract.cfm?URI=ao-53-20-4468},
doi = {10.1364/AO.53.004468},
}

@article{SapphireToTitanium,
    author = {Little, Bethany J. and Hoth, Gregory W. and Christensen, Justin and Walker, Chuck and De Smet, Dennis J. and Biedermann, Grant W. and Lee, Jongmin and Schwindt, Peter D. D.},
    title = {A passively pumped vacuum package sustaining cold atoms for more than 200 days},
    journal = {AVS Quantum Science},
    volume = {3},
    number = {3},
    pages = {035001},
    year = {2021},
    month = {07},
    issn = {2639-0213},
    doi = {10.1116/5.0053885},
    url = {https://doi.org/10.1116/5.0053885}
}

@Article{Lee2022,
author={Lee, Jongmin
and Ding, Roger
and Christensen, Justin
and Rosenthal, Randy R.
and Ison, Aaron
and Gillund, Daniel P.
and Bossert, David
and Fuerschbach, Kyle H.
and Kindel, William
and Finnegan, Patrick S.
and Wendt, Joel R.
and Gehl, Michael
and Kodigala, Ashok
and McGuinness, Hayden
and Walker, Charles A.
and Kemme, Shanalyn A.
and Lentine, Anthony
and Biedermann, Grant
and Schwindt, Peter D. D.},
title={A compact cold-atom interferometer with a high data-rate grating magneto-optical trap and a photonic-integrated-circuit-compatible laser system},
journal={Nature Communications},
year={2022},
month={Sep},
day={01},
volume={13},
number={1},
pages={5131},
issn={2041-1723},
doi={10.1038/s41467-022-31410-4},
url={https://doi.org/10.1038/s41467-022-31410-4}
}

@article{COOPER2021101898,
  title   = {Additively manufactured ultra-high vacuum chamber for portable quantum technologies},
  journal = {Additive Manufacturing},
  volume  = {40},
  pages   = {101898},
  year    = {2021},
  issn    = {2214-8604},
  doi     = {10.1016/j.addma.2021.101898},
  url     = {https://www.sciencedirect.com/science/article/pii/S2214860421000634},
  author  = {N. Cooper and L.A. Coles and S. Everton and I. Maskery and R.P. Campion and S. Madkhaly and C. Morley and J. O'Shea and W. Evans and R. Saint and P. Kr{\"u}ger and F. Oru{\v c}evi{\'c} and C. Tuck and R.D. Wildman and T.M. Fromhold and L. Hackerm{\"u}ller}
}

@article{Test2021,
  author  = {Burrow, Oliver S. and Osborn, Paul F. and Boughton, Edward and Mirando, Francesco and Burt, David P. and Griffin, Paul F. and Arnold, Aidan S. and Riis, Erling},
  title   = {Stand-alone vacuum cell for compact ultracold quantum technologies},
  journal = {Applied Physics Letters},
  volume  = {119},
  number  = {12},
  pages   = {124002},
  year    = {2021},
  month   = {Sep},
  issn    = {0003-6951},
  doi     = {10.1063/5.0061010},
  url     = {https://doi.org/10.1063/5.0061010}
}

\end{document}